\documentclass[prl,aps,twocolumn]{revtex4-1}

\usepackage[pdftex]{graphicx}
\usepackage{amsbsy,amssymb,amsmath,bm,mathtools}

\usepackage{xspace}
\usepackage{bm}% bold math
\usepackage{color}

\usepackage{url}
\usepackage[section]{placeins}
\usepackage[colorlinks=true,linkcolor=blue,citecolor=blue]{hyperref}

\begin{document}

%\preprint{Preprint}

\title{Machine learning dynamics of phase separation in correlated electron magnets}

\author{Puhan Zhang}

\author{Preetha Saha}

\author{Gia-Wei Chern}
\affiliation{Department of Physics, University of Virginia, Charlottesville, VA 22904, USA}

\date{\today}

\begin{abstract}
We demonstrate machine-learning enabled large-scale dynamical simulations of electronic phase separation in double-exchange system. This model, also known as the ferromagnetic Kondo lattice model, is believed to be relevant for the colossal magnetoresistance phenomenon.  Real-space simulations of such inhomogeneous states with exchange forces computed from the electron Hamiltonian can be prohibitively expensive for large systems. Here we show that linear-scaling exchange field computation can be achieved using neural networks trained by datasets from exact calculation on small lattices. Our Landau-Lifshitz dynamics simulations based on machine-learning potentials nicely reproduce not only the nonequilibrium relaxation process, but also correlation functions that agree quantitatively with exact simulations. Our work paves the way for large-scale dynamical simulations of correlated electron systems using machine-learning models. 
\end{abstract}

\maketitle

Electronic phase separation is a salient characteristic of strongly correlated electron systems. Indeed, early studies have shown that a generic feature of doped Mott insulators is the strong tendency toward phase separation in which the doped holes are expelled from the antiferromagnetic domains~\cite{visscher74,schulz89,emery90,white00,yee15}. For example, the emergence of charge stripes in cuprate superconductors originates from the segregation of doped holes into one-dimensional metallic stripes embedded in a correlated insulating matrix~\cite{tranquada95,kivelson98,kivelson03,emery99}. Another prominent example is the complex textures observed in several manganites that exhibit the colossal magnetoresistance (CMR) effect~\cite{dagotto_book,dagotto05,moreo99,dagotto01,mathur03}. These nanoscale patterns are a mixture of metallic ferromagnetic clusters and insulating antiferromagnetic domains~\cite{fath99,renner02,salamon01}. An intriguing scenario of CMR is the field-induced percolating transition of metallic nano-clusters in such a mixed-phase state~\cite{uehara99,zhang02}.

Metastable states with complex patterns have also been observed in countless nonlinear and nonequilibrium processes~\cite{seul95,gollub99,cross12}. Several numerical techniques, such as Monte Carlo and molecular dynamics simulations, have been developed to study the phase separation and pattern formation in classical systems~\cite{gunton83,puri09}. For large-scale simulations, the phase-field model has proven an efficient and flexible approach to modeling the dynamics of complex textures during phase transitions~\cite{collins86,valls90,steinbach13}. While such empirical methods capture some universal features of phase transition dynamics, however, they are rather limited in their predictive power, especially for phase-separation phenomena of electronic origins. In particular, knowledge of the interplay between microscopic electronic processes and the mesoscopic spatial-temporal dynamics is often required for modern materials-by-design approaches. On the other hand, full microscopic modeling of the phase-separation phenomena is often impossible due to the expensive electronic structure calculation or even many-body methods that are required at every time-step of the dynamical simulations.

Machine learning (ML) offers a promising solution to the difficult task of multi-scale modeling of electronic systems. Indeed, ML has been exploited for this purposes in {\em ab initio} molecular dynamics (MD) simulations for more than a decade~\cite{behler07,bartok10,li15,behler16,smith17,zhang18,deringer19,mueller20}.  The central idea of this approach is to develop deep-learning neural networks that emulate the time-consuming density function theory (DFT) calculations required at every MD time-step. Importantly, the neural net trained from small-size datasets can enable accurate MD simulations at much larger scales. This success has spurred recent efforts to calculate the training data at levels of quantum theory significantly beyond DFT~\cite{mcgibbon17,chmiela17}, such as the Gutzwiller method for describing the correlation-induced metal-insulator transition~\cite{ma19,suwa19}.

In this paper, we show that ML can be used to develop linear-scaling effective spin potential that capture the microscopic physics of electronic phase separation. We demonstrate our approach on the double-exchange (DE) system~\cite{zener51,anderson55,degennes60}, which play a crucial role in our understanding of the CMR effect. The double-exchange model can also be viewed as a mean-field approximation of the magnetic phase of the Hubbard model~\cite{chern17,mukherjee14}, a canonical system of strong electron correlation. We show that neural network models, trained from datasets of small-size exact diagonalization, successfully reproduce the nonequilibrium dynamics of electronic phase separation in double-exchange system.

We consider the single-band double-exchange (DE) model on a square lattice,
\begin{eqnarray}
	\label{eq:H_DE}
	\mathcal{H} = -t \sum_{\langle ij \rangle} \left( {c}^{\dagger}_{i \alpha} {c}^{\;}_{j \alpha} + {\rm h.c.} \right)
	- J \sum_{i} \mathbf S_i \cdot c^{\dagger}_{i\alpha}  {\bm{\sigma}_{\alpha\beta}} c^{\;}_{i\beta}, \quad
\end{eqnarray}
where ${c}^\dagger_{i \alpha}/c_{i, \alpha}$ are creation/annihilation operators of electron with spin $\alpha = \uparrow, \downarrow$ at site $i$,  $ \langle ij \rangle$ indicates the nearest neighbors, $t$ is the electron hopping constant, $J$ is the local Hund's rule coupling between electron spin and local magnetic moment $\mathbf S_i$, which are assumed to be classical spins of length $S = 1$. Here repeated indices $\alpha, \beta$ imply summation. The square-lattice DE model has been extensively studied theoretically~\cite{yunoki98,dagotto98,chattopadhyay01}. Exactly at half-filling, the local spins develop a long-range N\'eel order in the $T = 0$ insulating ground state. Upon hole doping, interestingly, a mixed-phase state consisting of hole-rich ferromagnetic puddles embedded in the half-filled antiferromagnetic insulator emerges as a stable thermodynamic phase at strong coupling~\cite{yunoki98,dagotto98,chattopadhyay01}, a scenario similar to that of the Hubbard model.

The structure of the mixed-phase state is intimately related to the spin configurations. For example, the ferromagnetic clusters are stabilized by the kinetic energy gain of doped holes propagating in a sea of parallel spins. The kinetics of phase separation is thus controlled by the spin dynamics, which is governed by the stochastic Landau-Lifshitz-Gilbert (LLG) equation~\cite{brown63}
\begin{eqnarray}
	\label{eq:LLG}
	\frac{d \mathbf S_i}{dt}\ = - \mathbf S_i \times \left( \mathbf H_i + \bm\eta_i \right) + \alpha \mathbf S_i \times \left(\mathbf S_i \times \mathbf H_i \right),
\end{eqnarray}
where $\mathbf H_i$ is the torque, or local exchange field, acting on spin-$i$, $\bm\eta_i(t)$ is a stochastic local field of zero mean, and $\alpha$ is the Gilbert damping coefficient. The noise $\bm\eta_i$ is assumed to be a Gaussian random variables with variance determined from $\alpha$ and temperature $T$ by the fluctuation-dissipation theorem. 

In LLG simulations of the DE model, the calculation of the local torques $\mathbf H_i = -\partial E/\partial \mathbf S_i$, which is given by the derivative of total energy, has to be carried out at every time-step. For adiabatic spin dynamics, the energy is given by
\begin{eqnarray}
	\label{eq:E_system}
	E = \langle \mathcal{H} \rangle = {\rm Tr}(\rho \mathcal{H}),
\end{eqnarray}
where $\rho = \exp(-\mathcal{H}/k_B T)$ is the density matrix of the equilibrium electron liquid. However, Repeated calculation of $\rho$ based on, e.g. exact diagonalization, can be overwhelmingly time consuming. Similar computational complexity also occurs in Monte Carlo (MC) simulations of DE models, in which a proposed spin update requires computing the energy difference $\Delta E = {\rm Tr}(\Delta \rho\,\mathcal{H})$. Recently, ML techniques have been applied to speed up MC simulations of DE model and other related systems~\cite{huang17,liu17,liu17b}. In the so-called self-learning MC method, trained ML models, in the form of either effective spin Hamiltonian or restricted Boltzmann machine, are used to propose cluster updates for MC sampling. Standard Metropolis algorithm, combined with exact solution for computing the energy difference, is used to determine the acceptance of the proposed update. It has been demonstrated that the autocorrelation time is dramatically reduced in such self-learning MC algorithms~\cite{huang17,liu17,liu17b}.

%In the stochastic LLG simulation, the calculation of the local torques ${\mathbf H}_{i}$ requires solving the electron tight-binding Hamiltonian to obtain energy $E = \langle  {\mathcal{H}} \rangle$ at every time-step. This process can be overwhelmingly time consuming, usually scaling cubed with the problem dimension involving large size matrix diagonalization. Linear scaling methods such as kernel polynomial method (KPM)~\cite{furukawa04,alvarez05,weisse06,barros13,barros14,wang18} have been developed with a wide range of applications in physics. ML method, also linear scaled, offers a new alternative and can be generalized to more situations such as models including explicit electron-electron interactions as in Hubbard model.

\begin{figure}
\includegraphics[width=1.0\columnwidth]{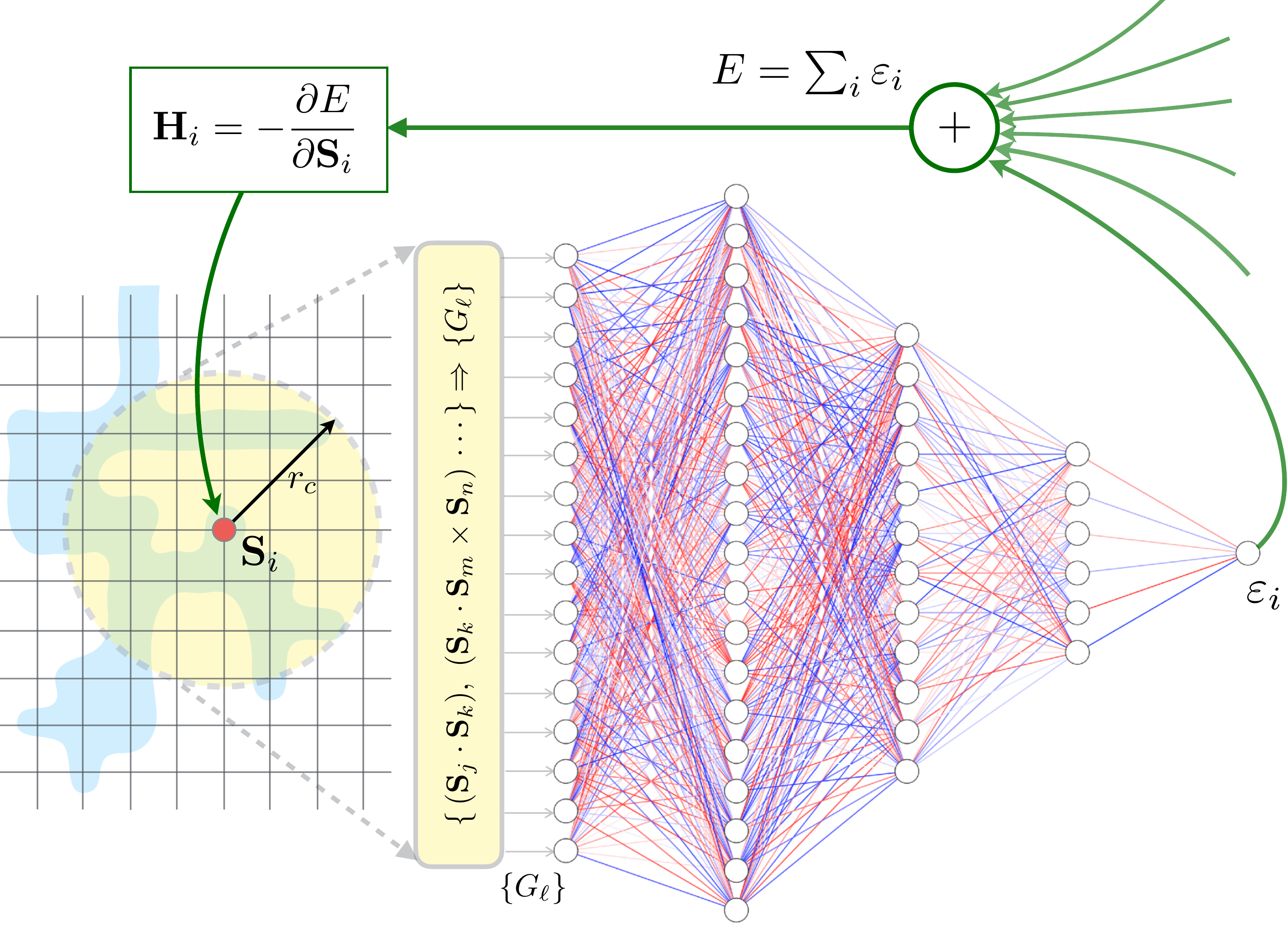}
\caption{Schematic diagram of neural-network (NN) potential model for LLG dynamics simulation of DE system. A descriptor is used to construct effective coordinates $\{G_\ell\}$, which are input to the NN, from neighborhood spins $\{\mathbf S_j\}$ up to a cutoff~$r_c$. The output of the NN corresponds to the local spin energy~$\varepsilon_i$. The total energy is obtained from output from all spins in the lattice, and its vector derivative with respective to spins gives the torques or local exchange fields $\mathbf H_i$.}
\label{fig:ml-scheme}  \
\end{figure}

%Linear scaling methods, such as the kernel polynomial method~\cite{furukawa04,alvarez05,weisse06,barros13,barros14,wang18}, have proven efficient for solving large-dimension Hamiltonian matrices in DE systems.

%ML offers alternative linear-scaling method, self-learning MC. Our ML approach, however, is in the same spirit as the ML-MD.

%Our ML approach offers an alternative linear-scaling method for adiabatic dynamics of DE models.  

Our ML approach to DE-type systems, outlined in Fig.~\ref{fig:ml-scheme}, is completely different from the above self-learning method and goes beyond the MC framework. In fact, our ML model is similar in spirit to linear-scaling ML potentials for large-scale MD simulations with quantum accuracy. In this scheme, the total energy $E$ in Eq.~(\ref{eq:E_system}) is expressed as a sum of local spin energies
\begin{eqnarray}
	E = \sum_i \varepsilon_i(\{\mathbf S_j\}),
\end{eqnarray}
where the energy $\varepsilon_i$ associated with the $i$-th spin is determined by its local environment within a cutoff radius~$r_c$. The partition of $E$ into local energies, called  atomic energies in the MD literature, is the starting point of most ML interatomic potentials~\cite{behler16,deringer19,mueller20}. The validity of this decomposition relies on the locality, or the rapid decay of correlation functions, of the electronic system.  While this approach is validated by our results to be discussed below, we have also explicitly checked the localization of spin energy from exact calculation of the DE model. 

Importantly, in our approach, the dependence of $\varepsilon_i$ on the neighborhood spin configuration $\{\mathbf S_j\}$ is encoded in a feed forward neural network (NN), which can be trained from exact solutions on small lattices. To ensure that symmetry of the DE Hamiltonian is preserved in the local spin energy function $\varepsilon_i(\{\mathbf S_j\})$, we have developed a descriptor that translates local spin environment into effective coordinates $\{G_\ell\}$ that are invariant under relevant symmetry operations.  The DE Hamiltonian in Eq.~(\ref{eq:H_DE}) is invariant under two independent symmetry groups: the SO(3) rotation of spins and the D$_4$ point group of the square lattice. The SO(3) rotation symmetry can be manifestly maintained by using only bond variables $b_{jk}$ and scalar chirality $\chi_{jmn}$: 
\begin{eqnarray}
	b_{jk} = \mathbf S_j \cdot \mathbf S_k, \qquad \chi_{jmn} = \mathbf S_j \cdot \mathbf S_m \times \mathbf S_n,
\end{eqnarray}
as building blocks for the effective coordinates. The collection of these variables around the $i$-th spin $\{b_{jk}, \chi_{jmn} \}$~form a high-dimensional representation of the D$_4$ group, which is then decomposed into the fundamental irreducible representations (irrep). The coefficients of each irreps $f^{A_1}_r$, $f^{A_2}_r$, $\cdots$, $\bm f^E_r$, where $r$ enumerates the multiplicity of each irrep, are proper linear combinations of the bond and scalar chirality variables. Finally, the generalized coordinates that are invariant under both symmetry operations are obtained from the amplitudes and relative phases of these irrep coefficients~\cite{ma19}. The various steps of our descriptor is summarized here
\begin{eqnarray}
	\{\mathbf S_j\} \,\, \to \,\, \{b_{jk}, \ \chi_{jmn} \} \,\, \to \,\, \{f^{\rm irrep}_r\} \,\, \to \,\, \{G_\ell\}. \nonumber
\end{eqnarray}
The resultant feature variables $\{G_\ell\}$ are input to the NN. Importantly, the local energy $\varepsilon_i(\{G_\ell\})$ obtained from the NN model depends only on the effective coordinates, hence obviously preserving the Hamiltonian symmetry.

\begin{figure}
\includegraphics[width=1.0\columnwidth]{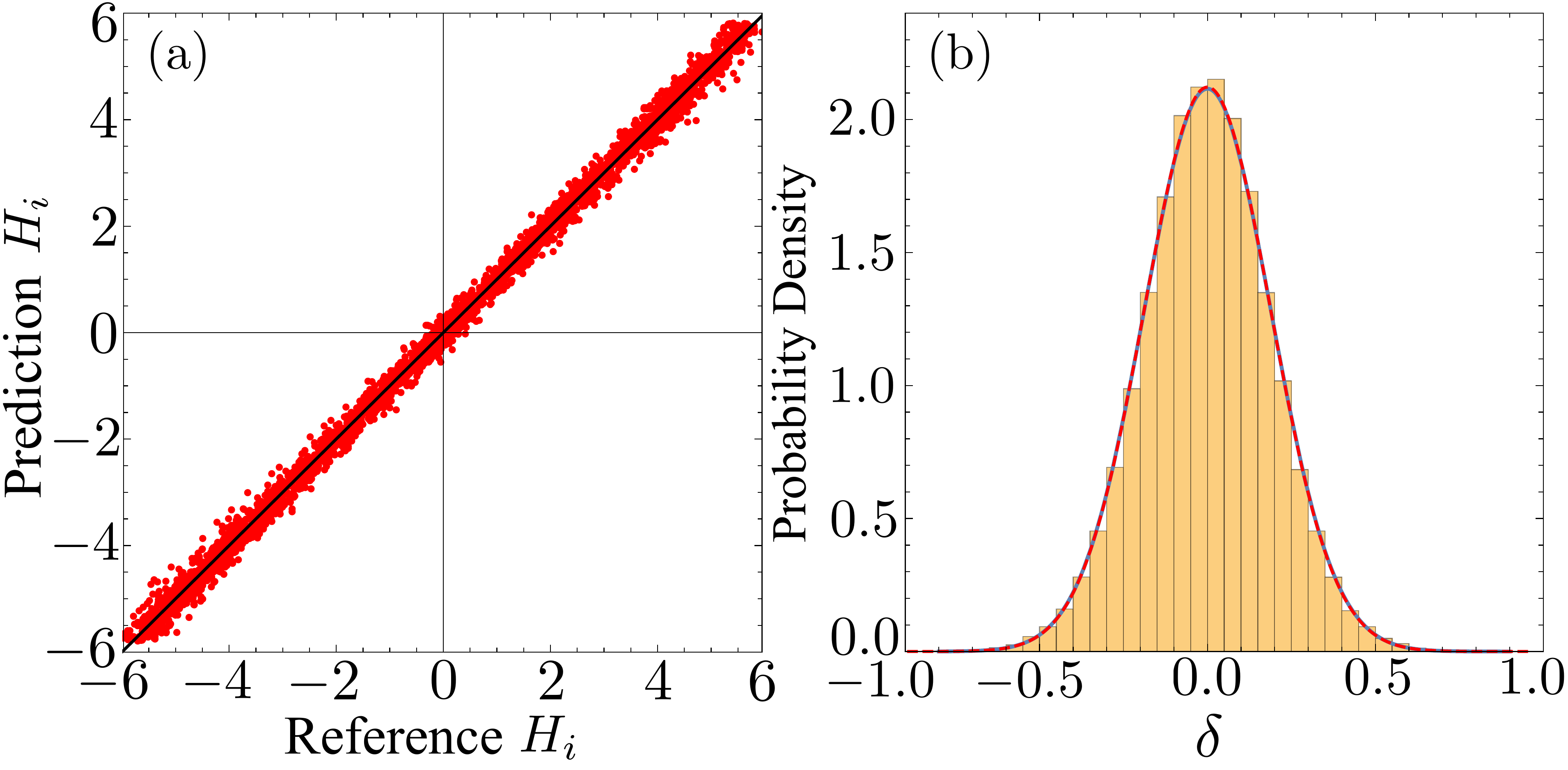}
\caption{Correlation between ML predicted torques and exact solutions from a test dataset. The cutoff of the ML descriptor is set at $r_c = \sqrt{26}$, including a total of 88 spins up to 14th nearest neighboring sites. Panel (b) shows the distribution of the force difference $\delta = H_{\text{ML}}-H_{\text{exact}}$ between ML prediction and exact diagonalization, which is well approximated by a normal distribution, shown as the red line, with a variance~$\sigma^2 = 0.035$. 
\label{fig:force} 
}
\end{figure}

An NN model for the spin energy is constructed and trained on PyTorch~\cite{paszke19}. The NN has 6 layers among which 4 of them are hidden layers consisting of $1024\times512\times256\times128$~\cite{nair10} neurons.  The input layer size is adjusted according to the number of feature variables~$G_{l}$. 
The NN performs a sequence of transformations on the input neurons, and a continuously differentiable exponential linear function~\cite{barron17} is used as the activation function between layers. The output layer consists of a single neuron whose value corresponds to the local energy $\varepsilon_i$. By combining outputs from all spins, automatic differentiation, as implemented in PyTorch~\cite{paszke17}, is employed to compute the derivative of total spin energy, $\partial E/\partial \mathbf S_i$, which gives rise to torques acting on spins; see Fig.~\ref{fig:ml-scheme}.

The training dataset consists of 3500 snapshots of spins and local torques, obtained from exact LLG simulations of a $30\times 30$ DE system using exact diagonalization. The electron filling fraction is set to $n=0.485$ and the temperature is fixed at $T=0.001$. Among the training data, 1000 snapshots are random spin configurations, while 1500 correspond to phase-separated state. The remaining 1000 snapshots come from the nonequilibrium intermediate states during the relaxation of the system. A~benchmark model is developed first using only random spin configuration. This model is then further improved by datasets with mixed intermediate and steady-state results. The loss function includes the mean square error (MSE) of both the torques and total energy. %Each batch of data is the features generated from one snapshot including 900 spins. 
The parameters of the neural net are Kaiming initialized~\cite{he15}. During the training, 5-fold cross-validation is performed. Adam optimizer~\cite{kingma14} with an adaptive learning rate is applied for this process. 
%We also include an early stopping mechanism that functions when the training error and validation error cannot be further reduced. 
%Initially, we use the random spin configurations to train a benchmark model before we mix all data along the relaxation path for further training in order to achieve state-of-the-art results. 

\begin{figure}
\includegraphics[width=0.98\columnwidth]{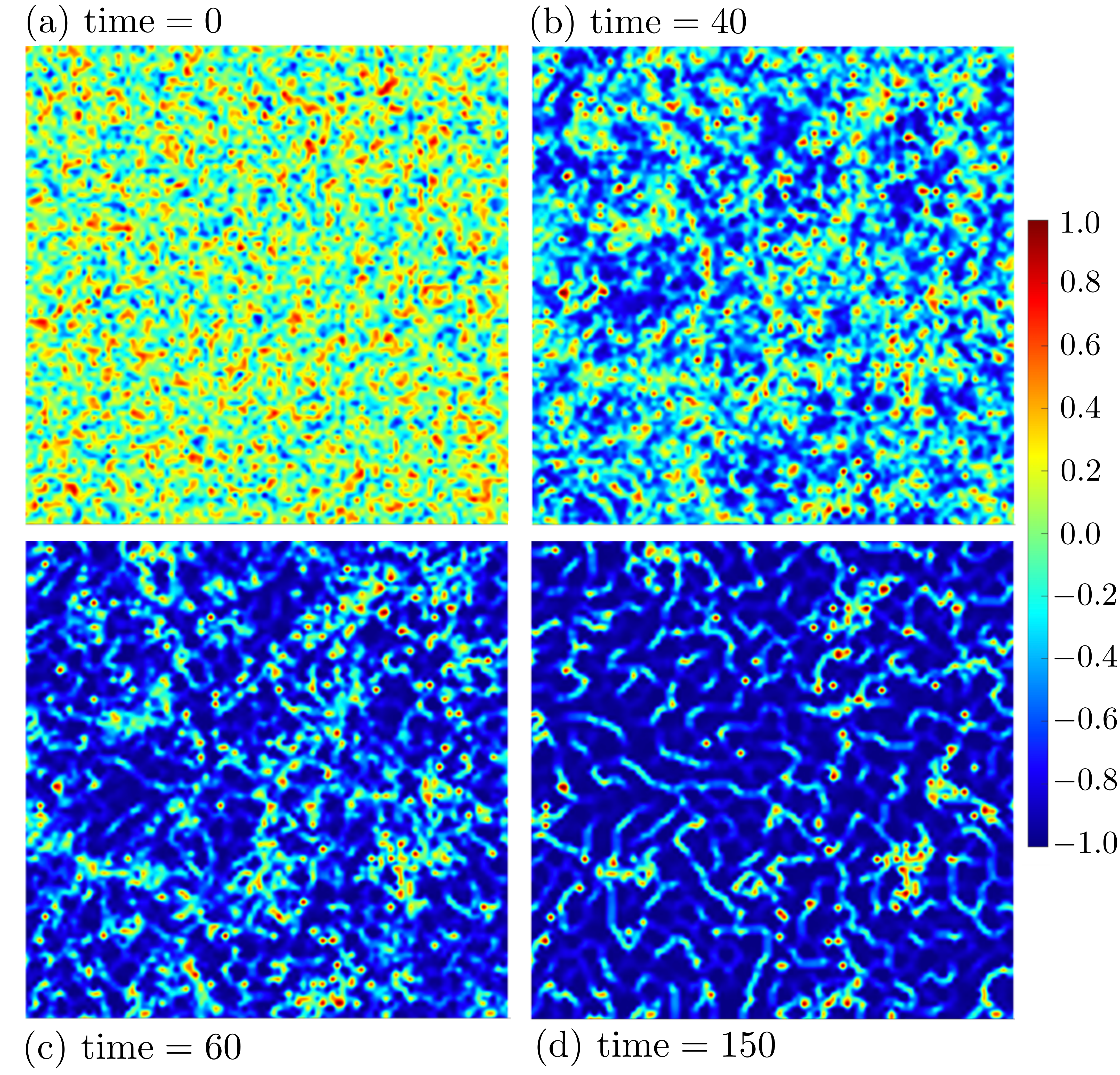}
\caption{Density plots of local nearest-neighbor spin correlation $C(\mathbf r_i) \equiv (\mathbf S_i \cdot \mathbf S_{i+\mathbf x} + \mathbf S_i \cdot \mathbf S_{i - \mathbf x} + \mathbf S_i \cdot \mathbf S_{i+\mathbf y} + \mathbf S_i \cdot \mathbf S_{i - \mathbf y})/4$ at four different simulation times of the ML-LLG dynamics simulation on a $100 \times 100$ lattice. The NN model is trained by datasets from exact solutions on a $L=30$ lattice, with parameters $t = 1$, $J = 7$, and electron filling fraction $n = 0.485$. The simulation time is measured in units of $t^{-1}$. 
\label{fig:relaxation} 
}
\end{figure}

Fig.~\ref{fig:force}(a) shows components of local torques $\mathbf H_i$ predicted by our trained NN model versus the exact results on a test dataset consisting of 500 snapshots of spins during the relaxation process. The difference $\delta = H_{{\text {ML}}} - H_{{\text {exact}}}$ between the ML prediction and exact calculation is well described by a Gaussian distribution, as shown in Fig.~\ref{fig:force}(b). A rather small MSE of $\sigma^2 = 0.035$, derived by variance of the Gaussian distribution indicates the high accuracy of the NN potential. Interestingly, the normal distribution of the deviation $\delta$ implies that the statistical error of the ML model can be interpreted as an effective temperature in the Langevin-type dynamics.

We next integrate our NN model with the LLG solver to perform large-scale dynamical simulations of the DE model. Fig.~\ref{fig:relaxation} shows density plots of the local spin-correlation obtained by averaging over 4 nearest-neighbor bonds, $C_i \equiv (\mathbf S_i \cdot \mathbf S_{i+\mathbf x} + \mathbf S_i \cdot \mathbf S_{i - \mathbf x} + \mathbf S_i \cdot \mathbf S_{i+\mathbf y} + \mathbf S_i \cdot \mathbf S_{i - \mathbf y})/4$, at different times during the relaxation of an initially random spin configuration. The NN model used in the LLG simulation was trained by datasets on a small $L = 30$ lattice and 1.5\% doped holes.  Regions with positive (negative)~$b_i $ in Fig.~\ref{fig:relaxation} correspond to predominately antiferromagnetic (ferromagnetic) alignment of spins. Our NN model clearly reproduces the nonequilibrium relaxation towards the phase separated state: the development of large domains of antiparallel spins interspersed with small ferromagnetic clusters. Moreover, we have used the kernel polynomial method~\cite{weisse06,barros13,wang18} to explicitly verified that the doped holes are confined to the ferromagnetic puddles, consistent with previous studies~\cite{jing19}.
Interestingly, in addition to aggregation around FM clusters, a fraction of the doped holes are self-trapped in a composite object, indicated by bright dots in Fig.~\ref{fig:relaxation}(d), which are magnetic analog of the polaron~\cite{jing19}.

\begin{figure}
\includegraphics[width=0.98\columnwidth]{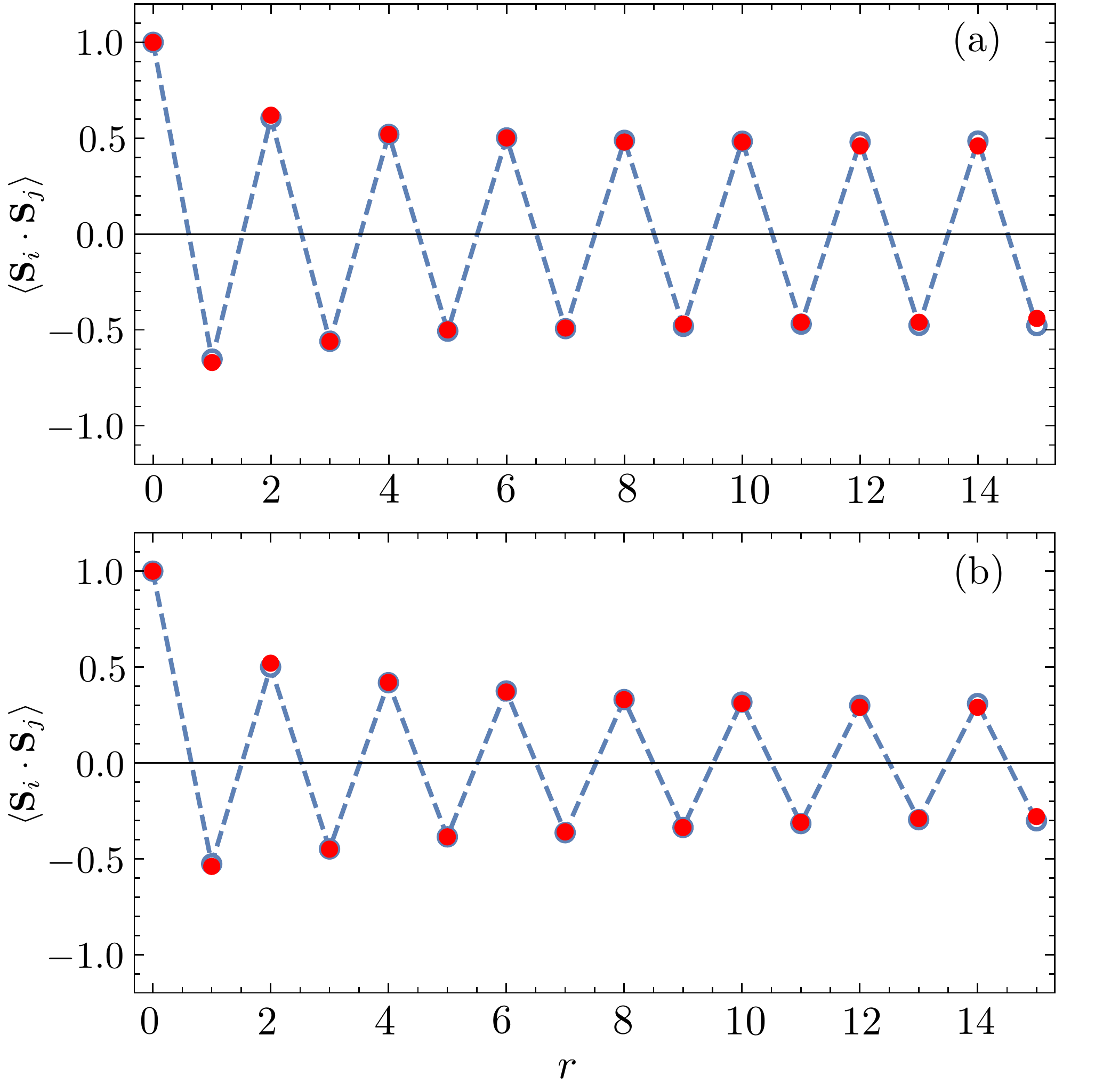}
\caption{Correlation function $C_{ij} = \langle \mathbf S_i \cdot \mathbf S_j \rangle$ as a function of $r_{ij} = |\mathbf r_j - \mathbf r_i|$ along the $x$ direction at electron filling fraction (a) $n = 0.485$ and (b) $n = 0.475$. The results were obtained by averaging over all lattice sites and tens of independent runs. The red dots are results from LLG simulations with NN models, while the blue dashed lines correspond to exact-diagonalization LLG simulations at~$T = 0.022$.  
\label{fig:corr}
}
\end{figure}

The spin-spin correlation function $C_{ij} = \langle \mathbf S_i \cdot \mathbf S_j \rangle$ obtained from the ML-LLG simulations is shown in Fig.~\ref{fig:corr} for two different electron filling frations. The predominately antiferromagnetic background results in a staggering of the spin correlation $C(x, y) \sim (-1)^{x+y}\, A(x, y)$, while the ferromagnetic clusters of variable sizes and randomly embedded in the N\'eel background cause a decay of the amplitude $A$. Although the ML-LLG simulations were performed without thermal noise, interestingly, the resultant correlation functions agree very well with those of the exact LLG simulations obtained at temperature $T = 0.022$. This thermal-like fluctuation is due to the intrinsic statistic error of the ML model. Moreover, since each torque prediction of the neural net is independent of each other, the noises introduced at different sites or different time-steps are uncorrelated. These observations thus suggest that the normal distributed prediction errors of the NN model can be modeled as an effective temperature in a Langevin-type simulations.

It is worth noting that the electronic phase separation shown in Fig.~\ref{fig:relaxation} relies on the coexistence of hole-rich ferromagnetic cluster and the half-filled N\'eel domain. Microscopically, this coexistence originates from the non-local spin-spin interactions mediated by itinerant electrons. However, stabilization of such inhomogeneous noncoplanar spin structures is highly nontrivial from the viewpoint of effective classical spin Hamiltonian. As demonstrated in previous studies~\cite{akagi12, momoi97,martin08,nandkishore12}, noncoplanar spin order often requires either multi-spin interaction such as $ \sum b_{ij}\, b_{kl}$, or scalar chirality terms $\sum \chi_{ijk} \chi_{mnl}$ in the Hamiltonian. This means that the local energy of spin-$i$ must include off-center bonds $b_{kl}$ with $k, l \neq i$, and scalar chirality $\chi_{ijk}$. Indeed, we found that NN models without such terms exhibit a much larger prediction error and cannot  reproduce the phase-separated states in LLG simulations. It is also remarkable that when all such terms are properly included in the input, the multi-layer NN automatically ``learn" the structure of the multi-spin interactions without human intervention. 

To summarize, we have successfully demonstrated ML-enabled large-scale simulations of phase separation phenomena in DE systems. Our ML-LLG dynamics framework is similar in spirit to those MD simulations based on ML interatomic potentials, such as the Behler-Parrinello scheme. A descriptor is developed to generate effective coordinates  that are invariant under both continuous spin-rotation and discrete lattice symmetries  from local spin environment.  Our deep-learning NN models, trained by exact quantum calculations on small systems, capture the highly frustrated spin-spin interactions that are essential to the inhomogeneous noncoplanar spin structures underlying the phase-separated states. 

Our work opens new avenues for using deep-learning models to simulate and understand large-scale dynamical phenomena of correlated lattice systems. For example, although as a proof of principle, we focus on the single-band double-exchange Hamiltonian, which is an oversimplified model for the CMR phenomena in manganites, our descriptor and NN model can be directly generalized to more realistic Hamiltonians which include multiple orbitals and Jahn-Teller effect. Moreover, long-range Coulomb interaction, which also plays an important role in phase separation and CMR phenomena, can be straightforwardly integrated with our ML models. More broadly, we envision our ML framework, when combined with more sophisticated many-body techniques such as dynamical mean-field theory or quantum Monte Carlo, could enable large-scale dynamical simulations of phase separation in Hubbard-type models.

{\em Acknowledgements}. The authors thank K. Barros and Sheng Zhang for useful discussions. The work was supported by the US Department of Energy Basic Energy Sciences under Contract No. DE-SC0020330. The authors also acknowledge the support of Advanced Research Computing Services at the University of Virginia.

\end{document}